\documentclass[aps,pra,preprint,amsmath,amsfonts,superscriptaddress,nofootinbib]{revtex4-1}

\newcommand{\R}{\mathbb{R}}
\newcommand{\Z}{\mathbb{Z}}
\newcommand{\into}{\hookrightarrow}

\usepackage[T1]{fontenc}
\usepackage[utf8]{inputenc}
\usepackage{braket}
\usepackage{bm}
\usepackage{graphicx,amsmath,revsymb,amssymb,array,bm,adjustbox,multirow,soul,changes,changebar,natbib}

\begin{document}

\title{Anyons from Three-Body Hard-Core Interactions in One Dimension}

\author{N.L. Harshman}
\affiliation{Department of Physics, American University, Washington, DC, USA}

\author{A.C. Knapp}
\affiliation{Department of Mathematics and Statistics, American University, Washington, DC, USA}

\date{\today}

\begin{abstract}
Traditional anyons in two dimensions have generalized exchange statistics governed by the braid group. By analyzing the topology of configuration space, we discover that an alternate generalization of the symmetric group governs particle exchanges when there are hard-core three-body interactions in one-dimension. We call this new exchange symmetry the traid group and demonstrate that it has abelian and non-abelian representations that are neither bosonic nor fermionic, and which also transform differently under particle exchanges than braid group anyons. We show that generalized exchange statistics occur because, like hard-core two-body interactions in two dimensions, hard-core three-body interactions in one dimension create defects with co-dimension two that make configuration space no longer simply-connected. Ultracold atoms in effectively one-dimensional optical traps provide a possible implementation for this alternate manifestation of anyonic physics.
\end{abstract}

\maketitle

\section{Introduction}

Particle exchange statistics are normally described by the symmetric group $S_N$ of particle permutations and indistinguishable particles are classified as bosons or fermions. However, there are more exotic possibilities for particle exchange statistics in low-dimensional particle models with hard-core interactions. The most famous examples are: 1) hard-core two-body interactions in one dimension, which leads to the `fermionization' of hard-core bosons~\cite{girardeau_relationship_1960}; and 2) hard-core two-body interactions in two dimensions, which allows for particles that are neither fermions nor bosons called anyons~\cite{wilczek_quantum_1982}.

Both of these manifestations of generalized exchange statistics can be understood topologically~\cite{leinaas_theory_1977}. The removal of two-body coincidences like ${\bf x}_i = {\bf x}_j$ from configuration space reduces its connectivity, and so does identifying points in configuration space that represent indistinguishable configurations.  In one dimension, the defects introduced by hard-core two-body interactions prevent particles from passing through each other. This divides configuration space into sectors with fixed order that are disconnected from each other. The relative phases among different sectors are unobservable, and therefore the exchange symmetry of indistinguishable particles is equivalent to a gauge symmetry~\cite{polychronakos_non-relativistic_1989}. In two dimensions, the defects caused by hard-core two-body interactions leave the configuration space connected but not simply-connected~\cite{wu_general_1984}. Exchanges of indistinguishable particles acquire different topological phases (or more general path-dependent transformations) depending on how they wind around these defects, leading to anyons with generalized exchange statistics governed by the fundamental group of configuration space, called the braid group $B_N$~\cite{wilczek_fractional_1990}.

Besides these two famous examples, there is only one other case where local hard-core interactions lead to a not simply-connected configuration space: hard-core three-body interactions in one dimension~\cite{harshman_coincidence_2018}. Consider the $N$-body Hamiltonian in one-dimensional free space with the form
\begin{equation}\label{eq:sampHam}
H =  \sum_{i=1}^N \left( -\frac{\hbar^2}{2m} \frac{\partial^2}{\partial x_i^2} + V(x_i) \right) + g \sum_{\langle ijk \rangle}^N W(\rho_{ijk}^2),
\end{equation}
where $V(x)$ is a finite one-body trapping potential, the second sum is over all triplets of particles $\langle ijk \rangle$, the  quadratic form $\rho_{ijk}^2=x^2_i +x^2_j + x^2_k-x_i x_j -x_j x_k -x_k x_i$ is the square of the three-body hyperradius for the triplet $\langle ijk \rangle$, and $W(\rho^2)$ is some repulsive three body potential with finite range. Since the range of the interaction will not affect topological properties, a mathematically convenient choice for $W(\rho_{ijk}^2)$ has support at  $\rho_{ijk}=0$ and nowhere else
\begin{equation}
W(\rho_{ijk}^2) = \delta(x_i-x_j)\delta(x_j-x_k).
\end{equation}
In the limit $g\to \infty$ in (\ref{eq:sampHam}), the three-body potential becomes hard-core and therefore the two-dimensional coincidence manifolds defined by $x_i = x_j = x_k$ are excluded from the $N$-dimensional  configuration space.

There are several proposals to create tunable effective three-body interactions in the control of ultracold atoms in optical traps~\cite{cooper_exact_2004, paredes_pfaffian-like_2007, daley_atomic_2009, keilmann_statistically_2011,mahmud_dynamically_2014,paul_large_2015,paul_hubbard_2016,strater_floquet_2016}. These proposals are driving sustained theoretical interest in the dynamical and thermodynamical properties of such models~\cite{silva-valencia_first_2011,paredes_non-abelian_2012,sowinski_exact_2012,safavi-naini_first-order_2012,wright_nonequilibrium_2014,hincapie-F_mott_2016,lange_strongly_2017,andersen_hybrid_2017,sekino_quantum_2018,drut_quantum_2018,nishida_universal_2018,arcila-forero_three-body-interaction_2018,pricoupenko_pure_2018,guijarro_one-dimensional_2018,pastukhov_ground-state_2019,pricoupenko_three-body_2019}. Two-body interactions could also be added to (\ref{eq:sampHam}) without affecting the connectivity, as long as they are not also hard-core.

\begin{figure}
\centering
\adjustbox{trim={.10\width} {.15\height} {0.10\width} {.12\height},clip}{\includegraphics[width= 1.2\columnwidth]{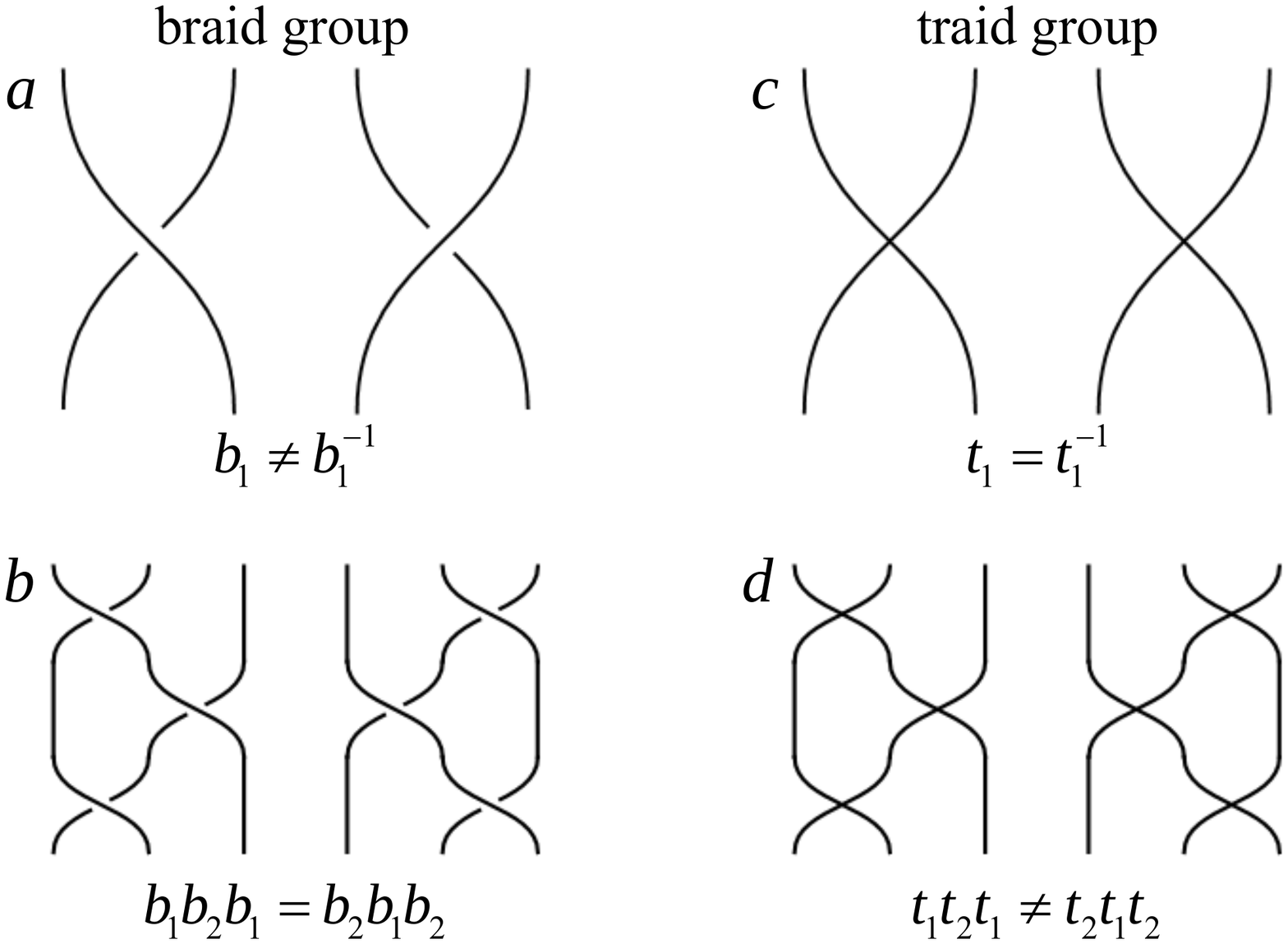}}
\caption{This figure compares the braid group $B_N$ and traid group $T_N$ using strand diagrams. Both groups can be expressed in terms $N-1$ generators, $b_i$ and $t_i$ respectively, that exchange the $i$th and the $(i\!+\!1)$th strands. For the braid group, $b_i$ and $b_i^{-1}$ are different, i.e.\ strands must go over or under each other, whereas for the traid group (like the symmetric group) two-particle exchanges $t_i^2=1$ are square-trivial and there is no distinction between over and under. Another contrast, for the traid group any rearrangement of strands that requires moving through a triple point is not allowed, whereas for the braid group shift one can pass through a triple point to show that the arrangements $b_i b_{i+1} b_i$ and $b_{i+1} b_i b_{i+1}$ of three adjacent strands are equivalent.}\label{fig:braidtraidcontrast}
\end{figure}

Hard-core three-body interactions in one dimension disrupt the connectivity of configuration space in a similar way to two-body coincidences in two-dimensions. In both cases the excluded coincidences create co-dimension $\tilde{d}= 2$ defects (see Tab.~\ref{tab:codimension}) around which paths realizing particle exchanges can wind and tangle. Because the configuration space is not simply connected, particle models with co-dimension $\tilde{d}= 2$ defects also possess anyonic solutions. Unlike fermionic and bosonic solutions, but similar to braid group anyons, these multi-valued solutions possess generalized exchange statistics realized by topological (or Berry) phases that depend on the path taken by the particle exchange. 
However, instead of obeying the familiar braid group exchange statistics, the anyonic solutions  of Hamiltonians like (\ref{eq:sampHam}) obey generalized winding rules described by a group we call the \emph{traid group}. Like the braid group, the traid group is an extension of the symmetric group and its elements can be represented as strand diagrams; see Fig.~\ref{fig:braidtraidcontrast}.

\begin{table}%
\centering
\begin{tabular}{c|ccc}
\hline
 & $k=2$ & $k=3$ & $k=4$  \\
\hline
$d=1$ & $1$ & $2$ & $3$\\
$d=2$ & $2$ & $4$ & $6$\\
$d=3$ & $3$ & $6$ & $9$\\
\hline
\end{tabular}
\caption{The entries of this table are the co-dimension $\tilde{d}$ of the defect created when $k$-body coincidences are removed from the configuration space of $N$ particles in $d$ dimensions. The formula for the co-dimension $\tilde{d}= d(k-1)$ counts the number of equations necessary to establish a $k$-body coincidence in $d$ dimensions and is therefore independent of $N$. When $\tilde{d}=1$, the defect is like a line in a plane or a plane in a space and it divides configuration space into dynamically isolated sectors. When $\tilde{d}=2$, the defect is like a point in a plane or a line in a space. Configuration space remains connected, but not simply-connected and topological phases are possible. When $\tilde{d}>2$, configuration space remains simply connected and generalized exchange statistics are not possible.}
\label{tab:codimension}
\end{table}

Although the original physical motivation for this study is engineered three-body interactions in ultracold atoms, real indistinguishable atoms are either bosons or fermions. For the anyonic solutions to be experimentally relevant, the one-dimensional particles of the model Hamiltonian (\ref{eq:sampHam}) would themselves need to be quasiparticles with an internal structure, perhaps at a faster time scale, which provides the effective path-dependent exchange phase of traid anyons. Although it seems unlikely that there are `natural' systems with low-energy dynamics given by a one-dimensional quasiparticle model with effective hard-core three-body interactions, the surprising applicability of braid anyons to the fractional quantum Hall effect provides inspiration for the exploration of this novel form of generalized exchange statistics. Even without physical instantiations of traid anyons, their mathematical structure possesses that delightful combination of simple to express but rich in expression that begs further investigation, if only as a clarifying contrast to the more famous exchange statistics given by the symmetric group and the braid group.

The article has the following structure:  In Sect.~\ref{sect:background}, we introduce notation and terminology for describing the homotopy of configuration space by briefly presenting the braid group and braid group anyons. Using these results, we introduce the traid group and and draw contrasts between traid anyons and previous results for braid anyons in one and two dimensions. Next, in Sect.~\ref{sect:config} we describe the geometrical and topological properties of configuration spaces for $N$ distinguishable and indistinguishable particles in $d$ dimensions with hard-core $k$-body interactions and analyze its topology in the case of hard-core three-body interactions in one dimension. One technical challenge, and an important difference with the braid group, is that the configuration space whose properties determine the generalized traid group exchange statistics is not a manifold. Instead it is an orbifold, and in Sect.~\ref{sect:fund} (and the Appendix) we explain this difference and the generalization of fundamental groups that applies to orbifolds. In Sect.~\ref{sect:pres} we give the abstract presentations of the traid group and pure traid group and compare them to the symmetric and braid groups. Some representations of the traid groups corresponding to anyonic solutions and an application to three harmonically-traped abelian traid anyons are given in Sect.~\ref{sect:repres}. Summary and outlook are provided in Sect.~\ref{sect:summ}.

\section{Background and Connections}\label{sect:background}

Generalized exchange statistics can occur when the configuration space of a particle model is not simply connected~\cite{leinaas_theory_1977,sudarshan_configuration_1988,polychronakos_non-relativistic_1989,einarsson_fractional_1990}. In generalized exchange statistics, the transformation of the $N$-body wave function depends not just on which particles were exchanged, but also on \emph{how} they were exchanged. Particle exchanges are considered as paths through configuration space, and when the configuration space is not simply connected, there are inequivalent paths representing the same particle exchange. The fundamental group of the configuration space describes the equivalence classes of exchange paths and the representations of the fundamental group determine whether generalized exchange statistics are possible.

In the most famous example of generalized exchange statistics, the group describing particle exchanges is the braid group $B_N$~\cite{artin_theorie_1925,arnold_cohomology_1969,birman_braids:_2005,kassel_braid_2008}. The braid group $B_N$ is an infinite, discrete group that generalizes the symmetric group $S_N$. Generalized exchange statistics obeying the braid group occur when two-body coincidences are excluded from the configuration space of $N$ indistinguishable particles in two-dimensional Euclidean space~\cite{wu_general_1984}. One reason to remove these points from configuration space is because hard-core two-body interactions exclude those coincidences. Alternatively, whenever particles have relative angular momentum in two dimensions, the $1/r^2$ singularity of the centrifugal barrier also prevents two-body coincidences. Either way, the removed points form what we call the two-body \emph{coincidence structure} $\mathcal{V}_{N,2,2}$. This structure is the union of $N(N-1)/2$ two-body coincidence manifolds $\mathcal{V}_{ij}$, one manifold for each pair of particles. Each manifold $\mathcal{V}_{ij}$ is a hyperplane with co-dimension $\tilde{d}=2$, i.e.\ there are two dimensions perpendicular to it, like a point in $\mathbb{R}^2$ or a line in $\mathbb{R}^3$. Because the two-body coincidence manifolds $\mathcal{V}_{ij}$ are co-dimension $\tilde{d}=2$ defects, they disrupt the simple connectivity of configuration space, allowing the wave function to get `wound up' when particles exchange.

After the two-body coincidence structure $\mathcal{V}_{N,2,2}$ is removed, the remaining configuration space for $N$ indistinguishable particles is $\mathcal{X}_{N,2,2}/{\mathrm{S}_N}$, where $\mathcal{X}_{N,2,2}=\mathbb{R}^{2N} - \mathcal{V}_{N,2,2}$ is the configuration space of $N$ distinguishable particles. In the quotient space $\mathcal{X}_{N,2,2}/\mathrm{S}_N$, all points in $\mathcal{X}_{N,2,2}$ that differ only by a permutation of particle coordinates are identified as the same point. 
The fundamental group $\pi_1(\mathcal{X}_{N,2,2}/S_N)$ is the braid group $B_N$ and describes the generalized exchange statistics of indistinguishable particles, and the fundamental group  $\pi_1(\mathcal{X}_{N,2,2})$ is a subgroup called the pure braid group $PB_N$ and describes the generalized exchange statistics of distinguishable particles. The one-dimensional representations of the braid group have fractional exchange statistics governed  by a phase that varies from $\theta=0$ for bosons to $\theta=\pi$ for fermions. Quasiparticles obeying abelian braid statistics are central to the understanding of the fractional quantum Hall effect~~\cite{halperin_statistics_1984,arovas_fractional_1984, jain_theory_1990, wilczek_fractional_1990, canright_fractional_1990, viefers_ideal_1995, khare_fractional_1997}. Non-abelian anyons carry multi-dimensional representations of the braid group~\cite{moore_nonabelions_1991,wen_non-abelian_1991,greiter_paired_1992} and provide a model for quantum computing with topological error protection~\cite{kitaev_fault-tolerant_2003,nayak_non-abelian_2008}.

Besides two-body hard-core interactions in two dimensions, the only other case where local hard-core few-body interactions make configuration space not simply-connected and lead to anyonic physics is the much less studied case of three-body hard-core interactions in one dimension\footnote{There is also a non-local four-body interaction in one-dimension that creates co-dimension $\tilde{d}=2$ defects. This non-local hard-core interaction excludes formation of more than one pair so that the coincidence manifolds defined by two two-body coincidences like $x_i = x_j$ and $x_k = x_l$ are excluded from configuration space. Although not discussed in this article, this also leads to novel generalized exchange statistics distinct from those given by the braid or traid group. We refer to this group as the `fraid' group; Khovanov has named this the triplet group~\cite{khovanov_real_1996}.}. The simultaneous coincidence of three particles in one-dimension $x_i=x_j=x_k$ defines a linear subspace with co-dimension $\tilde{d}=2$, and the union of these is the three-body coincidence structure $\mathcal{V}_{N,1,3}$. When these forbidden coincidences are removed from the configuration space for $N$ particles in one-dimension, the remaining space $\mathcal{X}_{N,1,3} = \mathbb{R}^N - \mathcal{V}_{N,1,3}$ is not simply connected, nor (in a generalized sense described below) is the quotient  $\mathcal{X}_{N,1,3}/S_N$. Identifying the topological properties of $\mathcal{X}_{N,1,3}$ and $\mathcal{X}_{N,1,3}/S_N$ is the main technical result of this article. 

\begin{figure}
 \centering
 \includegraphics[scale=0.25, angle=90]{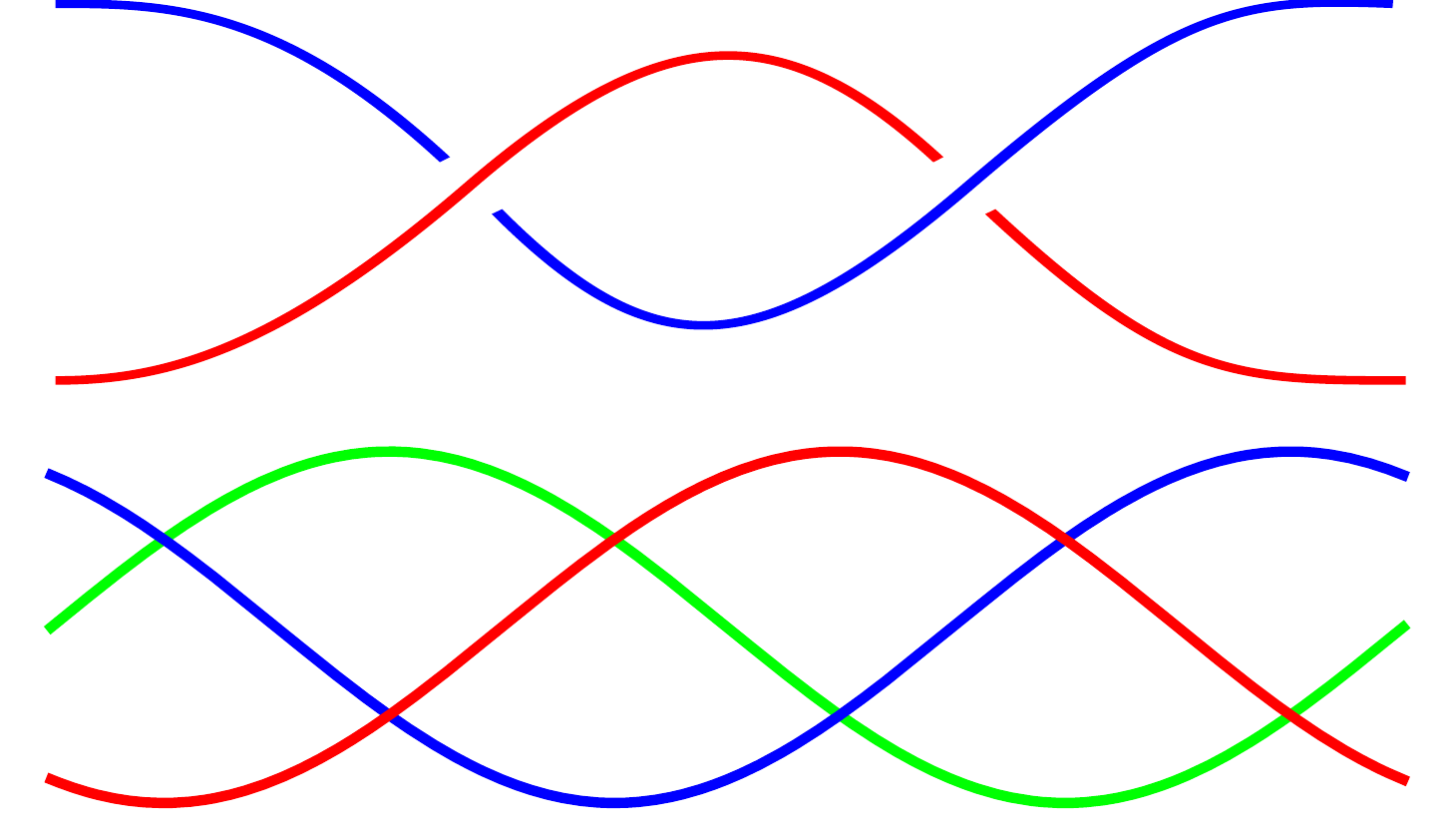}
 \caption{The pure braid group $PB_N$ and the pure traid group $PT_N$ apply to distinguishable particles and therefore group elements correspond to strand diagrams where all particles start and end at the same place. Here we depict the strand diagrams for the generators of $PB_2$ (top) and $PT_3$ (bottom). The configuration spaces for three particles with hard-core three-body interactions in one dimension
  $\mathcal{X}_{3,1,3}$ and for two particles with hard-core two-body interactions in two
  dimensions, $\mathcal{X}_{2,2,2}$ are homotopy equivalent to the circle $S^1$. Their fundamental groups $PT_3$ and $PB_2$ are both isomorphic to $\Z$ and generated by a choice of loop which starts at an arbitrary base point $x_0 \in S^1$ and travels around the circle once.} \label{fig:pure-traid-and-braid-choreography}
\end{figure}

In analogy with the pure braid group, we define the \emph{pure traid group} $PT_N$ as the fundamental group of the configuration space $\mathcal{X}_{N,1,3}$. We define the \emph{traid group} $T_N$ as the orbifold fundamental group \cite{thurston_geometry_2002} of $\mathcal{X}_{N,1,3}/S_N$. Like the braid groups, the traid groups are infinite non-abelian groups that have an intuitive diagrammatic representation in terms of weaving strands, see Figs.\ \ref{fig:braidtraidcontrast} and \ref{fig:pure-traid-and-braid-choreography}, but the weaving rules are different. Like the symmetric group, both the braid group $B_N$ and the traid group $T_N$ can be defined by the relations among $N-1$ generators corresponding to exchanging adjacent particles. We show below that $B_N$ and $T_N$ can be understood as two different ways of `loosening' $S_N$ symmetry: for $B_N$ the generators are no longer self-inverses; and for $T_N$ the generators no longer satisfy the Yang-Baxter relation (aka the braid relation or the third Reidemeister move). Breaking the Yang-Baxter relation allows for abelian and non-abelian representations of $T_N$ that exhibit generalized exchange statistics different from braid anyons. As we show below, $T_N$ is a linear hyperbolic Coxeter group~\cite{coxeter_regular_1973,humphreys_reflection_1992} with $N-1$ generators connected by infinitesimal angles, sometimes denoted $[\infty^{N-1}]$. The lowest traid group $T_3$ is isomorphic to the infinite dihedral group $D_\infty \sim [\infty]$ and the lowest pure traid group $PT_3$ (like $B_2$ and $PB_2$) is isomorphic to the group of integers.

In the large literature on anyons there are other generalizations of braid groups and their representations, e.g.~\cite{einarsson_fractional_1990,yamamoto_anyons_1994,read_non-abelian_2003,kauffman_virtual_2004,jacak_recovery_2010}, but the groups $PT_N$ and $T_N$ are distinct from any other groups analyzed in the context of generalized exchange statistics to the best of our knowledge. 
Relevant mathematical analysis was initiated by Bj\"orner and Welker~\cite{bjorner_homology_1995}. Motivated by the complexity theory of graphs, they analyzed the topology of configuration spaces for one-dimensional $\mathcal{X}_{N, 1, k}$ and two-dimensional systems $\mathcal{X}_{N, 2, k}$ with $k$-body coincidences removed. Subsequent work by mathematicians investigated the fundamental groups of these spaces and their quotients by the symmetric group. In this previous work, the traid groups $T_N$ and pure traid group $PT_N$ have been called the twin group and pure twin group~\cite{khovanov_real_1996, khovanov_doodle_1997, bardakov_structural_2019} or the planar braid group and pure planar braid group~\cite{gonzalez_linear_2019}.  
%One key difference between braid anyons and traid anyons is that in two-dimensions, particle exchanges take paths in space that enclose areas and the topological phase acquired by the wave function during that exchange can be associated (conceptually and mathematically) with a flux through that area. In contrast, the relevant exchange paths in one dimension do not encircle a flux in space, but encircle a flux in \emph{configuration space} which is higher dimensional and harder to visualize.
Interestingly, our results were anticipated by the earliest studies of the topological phase acquired in the adiabatic Born-Oppenheimer solutions of planar triatomic molecules~\cite{herzberg_intersection_1963, stone_spin-orbit_1976, mead_determination_1979}. Restrictions on the configuration space make this an effectively one-dimensional system with a singular three-body coincidence. As such, it provides an example of a system carrying a non-trivial representation of $PT_3$.

Despite occasional claims that anyons can only occur in two dimensions, there is a large body of previous work on one-dimensional anyons. This previous work has relied on one or more analogies to two-dimensional braid anyons, including: (1) obeying braid exchange statistics~\cite{polychronakos_non-relativistic_1989, ha_fractional_1995,zhu_topological_1996,kundu_exact_1999,girardeau_anyon-fermion_2006,hao_ground-state_2009, sree_ranjani_explicit_2009, keilmann_statistically_2011,wright_nonequilibrium_2014,zinner_strongly_2015,strater_floquet_2016,lange_strongly_2017}; (2) obeying generalized \emph{exclusion} statistics~\cite{haldane_``fractional_1991,haldane_``spinon_1991,ha_exact_1994, ha_fractional_1995, girardeau_anyon-fermion_2006,batchelor_one-dimensional_2006,patu_correlation_2007, vitoriano_fractional_2009}; or (3) having wave functions with either Laughlin (abelian braid anyons~\cite{laughlin_anomalous_1983}) or Pfaffian (non-abelian braid anyons~\cite{moore_nonabelions_1991,greiter_paired_1992}) forms~\cite{haldane_``spinon_1991,polychronakos_non-relativistic_1989,paredes_pfaffian-like_2007, girardeau_three-body_2010,paredes_non-abelian_2012}.

Note that all of these manifestations of braid anyon behavior appear naturally in the one-dimensional two-body hard-core Calogero-type models~\cite{polychronakos_non-relativistic_1989,ha_exact_1994,ha_fractional_1995, sree_ranjani_explicit_2009}.
One-dimensional models with fractional exchange statistics can also be constructed `by hand', i.e.\ inserting the phase $e^{i\theta}$ into the commutation relations of creation and annihilation operators~\cite{zhu_topological_1996}. But to emphasize, fractional exchange statistics are characteristic of abelian representations of the braid group and do not occur for the traid group, in which two-particle exchanges must be square-trivial. The abelian traid anyons we conjecture would obey a different form of generalized exchange statistics than the braid anyons.

Intriguingly, fractional exchange statistics occur `naturally' in some one-dimensional free-space models and lattice models with competing two-body \emph{and} three-body interactions, like the Kundu model~\cite{kundu_exact_1999} and the anyon-Hubbard model~\cite{paredes_pfaffian-like_2007, keilmann_statistically_2011, paredes_non-abelian_2012, wright_nonequilibrium_2014, strater_floquet_2016, lange_strongly_2017}. The Kundu model has highly-singular two-body and three-body interactions. For a certain balance of interaction strengths, the system can be transformed into an equivalent model with only two-body delta-interactions and `twisted' boundary conditions. This model is integrable and solvable by Bethe-ansatz~\cite{kundu_exact_1999,batchelor_one-dimensional_2006,patu_correlation_2007}. This is somewhat surprising, because generally three-body interactions break the Yang-Baxter relation (as in the traid group) and prevent integrability~\cite{ha_fractional_1995, sutherland_beautiful_2004}. In the anyon-Hubbard model, effective three-body interactions are created by occupation-number dependent hopping amplitudes~\cite{keilmann_statistically_2011,strater_floquet_2016} and the wave function has a Pfaffian form~\cite{paredes_pfaffian-like_2007,paredes_non-abelian_2012}. Solutions that have a Pfaffian from exhibit non-abelian braid statistics, and they can also be understood as the eigenstates of effective $k$-body hard-core interactions~\cite{greiter_paired_1991,read_beyond_1999,cooper_exact_2004,girardeau_three-body_2010}. 

However, no one has considered how the connectivity of the underlying configuration space for the Kundu or anyon-Hubbard models is disrupted by the co-dimension $\tilde{d}=2$ defects created by the addition of three-body interactions. Whether any of this previous work on one-dimensional anyons with three-body interactions can be reinterpreted in terms of traid groups remains an open question for future work. Additionally, implementing models with traid statistics in tight-binding lattice models with occupation-number dependent tunneling is an intriguing possibility.

\section{Configuration space}\label{sect:config}

The possibility for generalized exchange statistics is determined by the fundamental groups of two related spaces: (1) the configuration space $\mathcal{X}_{N,d,k}$ for $N$ distinguishable particles in $d$ dimensions with $k$-body hard-core interactions; and (2) the configuration space $\mathcal{X}_{N,d,k}/S_N$ for $N$ indistinguishable particles. When there are no interactions, the free configuration space of $N$ distinguishable particles in Euclidean space is the manifold $\R^{dN}=\left\{ (x_1,\ldots,x_N) \in \R^d\times \cdots \times \R^d \right\}$. For indistinguishable particles, the free configuration space is the orbifold $\mathcal{X}_{N,d,k}/S_N$, the quotient of the free configuration space by the symmetric group on $N$ objects. Configurations in $\R^{dN}$ that differ by only a permutation of particle positions are identified by the same point in $\R^{dN}/S_N$~\cite{leinaas_theory_1977,thurston_geometry_2002}.

The hard-core $k$-body interactions create impenetrable defects and complicate the topology of configuration space. Since the exact functional form of an interaction does not affect the connectivity, we can model the defects as zero-range, contact interactions without loss of generality. Then the defects are described by the coincidence structure $V_{N,d,k}$, defined as the union of all $\binom{N}{k}$ linear subspaces formed by coincidences of $k$ distinct particles $x_{i_1},\ldots,x_{i_k}$ in $d$-dimensions. Each linear subspace in the coincidence structure $V_{N,d,k}$ corresponds to $x_{i_1}=\cdots=x_{i_k}$ and has a co-dimension $\tilde{d}=d(k-1)$. Note that coincidence structures for higher-body interactions are nested inside lower-body interactions, i.e.\ when $k'>k$ then $V_{N,d,k'}\subset V_{N,d,k}$. The geometry and symmetries of these structures are analyzed in \cite{harshman_coincidence_2018}.

Removing the coincidence structure $V_{N,d,k}$ from the free configuration space $\R^{dN}$ gives the configuration space $\mathcal{X}_{N,d,k}= \R^{dN} - V_{N,d,k}$. Analyzing the connectivity of $\mathcal{X}_{N,d,k}$ is made simpler by using two symmetries to trivialize two degrees of freedom. First, $\mathcal{X}_{N,d,k}$ is invariant under translation in the $d$-dimensional linear subspace where all particles coincide $x_{1}=\cdots=x_{N}$, i.e.\ translations along the coincidence structure $V_{N,d,N} = \R^d$ corresponding to the the center-of-mass degree of freedom are a symmetry of $\mathcal{X}_{N,d,k}$. Second, the coincidence structure $V_{N,d,k}$ is constructed from linear subspaces that are scale invariant, and so the space $\mathcal{X}_{N,d,k}$ is also scale invariant. Combining these two symmetries, the configuration space can be factored into a reduced configuration space $\overline{\mathcal{X}}_{N,d,k}$ and two other terms $\mathcal{X}_{N,d,k} = \overline{\mathcal{X}}_{N,d,k} \times \R^+ \times \R^d$. Similarly, the free configuration space factors into $\R^{dN} =  S^{d(N-1)-1} \times \R^+ \times \R^d$.
Since $\R^+$ and $\R^d$ are homotopically trivial, $\mathcal{X}_{N,d,k}$ and $\overline{\mathcal{X}}_{N,d,k}$ are homotopy equivalent; i.e.\ they have the same connectivity and in particular they have isomorphic fundamental groups.

The reduced coincidence structure $\overline{V}_{N,d,k}$ can be defined by projection in the almost same way, as long as $V_{N,d,N}$ is added back in by hand $V_{N,d,k} = \left(\overline{V}_{N,d,k} \times \R^+ \times \R^d \right) \cup V_{N,d,N}$. For $N>k$ the space $\overline{V}_{N,d,k}$ is an arrangement of $\binom{N}{k}$ copies of spheres $S^{d(N-k)-1}$ contained inside the reduced free configuration space $S^{d(N-1)-1}$. These results agree with Thm.~1.3 of \cite{bjorner_homology_1995}, which analyzes the topology of $V_{N,1,k}$ and $V_{N,2,k}$ and proves they are homotopically equivalent to the wedge product of spheres.

\begin{figure}
 \centering
 \includegraphics[scale=0.2]{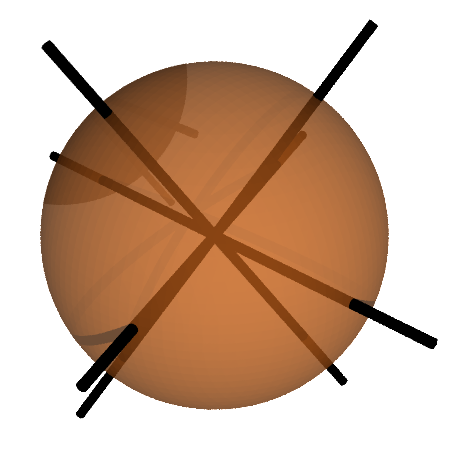}
 \includegraphics[scale=0.2]{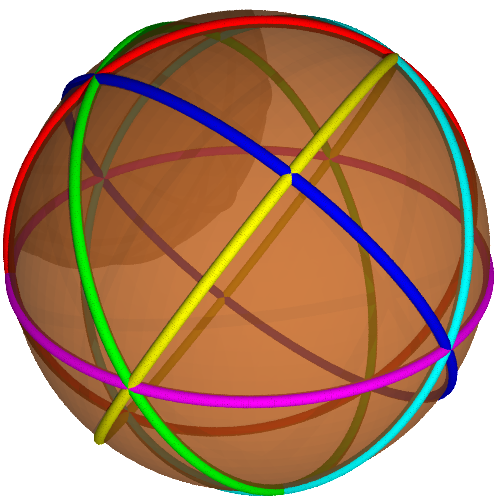}
 \includegraphics[scale=0.4]{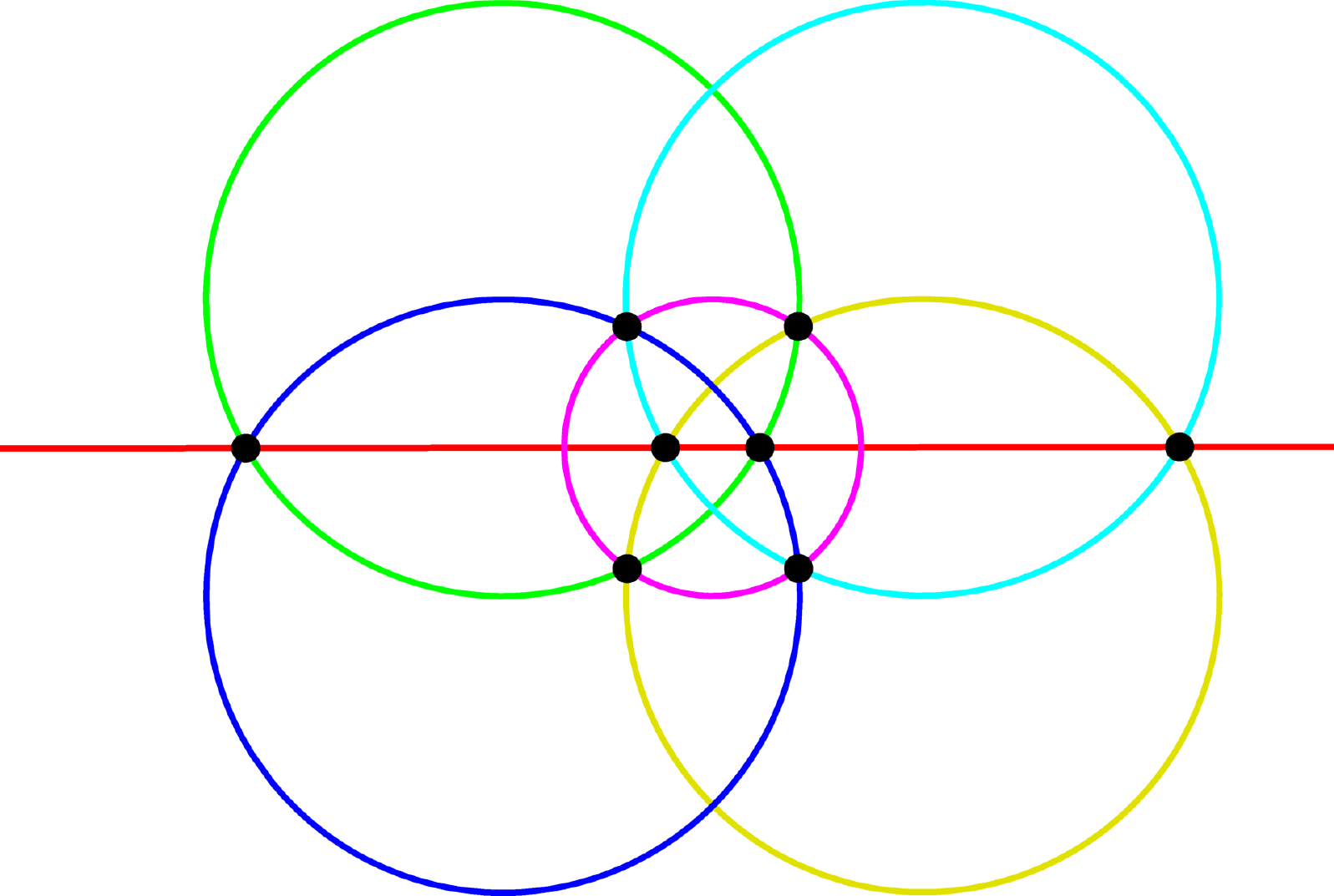}
 \caption{From top left, $\overline{\mathcal{X}}_{4,1,3}$ with $3$-coincidence loci as the intersection of $S^2$ with rays from the origin (black). Next, brown regions are $\overline{\mathcal{X}}_{4,1,2}$; formed by cutting $S^2$ along circles of $\overline{V}_{4,1,2}$. Finally, their mutual stereographic projection shows the $4!$ fundamental regions of the symmetric group action. Note that not all intersections of the $2$-coincidence locus correspond to $3$-coincidences; some correspond to commuting generators $t_1$ and $t_3$ of the traid group $T_4$.} \label{fig:reducedX413}
\end{figure}

As a relevant example, consider four particles in one dimension with hard-core two-body or three-body interactions. The spaces $\overline{\mathcal{X}}_{4,1,2}$ and $\overline{\mathcal{X}}_{4,1,3}$ are depicted in Fig.~\ref{fig:reducedX413}. The reduced free configuration space is the sphere $S^2$. The two-body reduced coincidence structure $\overline{V}_{4,1,2}$ is six intersecting great circles $S^1$ on the sphere. Because these circles have co-dimension $\tilde{d}=1$, they segment the reduced configuration space $\bar{\mathcal{X}}_{4,1,2}$ into 24 disconnected sectors corresponding to a specific orders of particles in one dimension. In the orbifold space $\overline{\mathcal{X}}_{4,1,2}/S_4$, these 24 sectors are identified with each other into a single sector representing configurations of four indistinguishable particles. The three-body reduced coincidence structure $\overline{V}_{4,1,3}$ is four copies of $S^0$ (i.e.\ two points) and the reduced configuration space $\overline{\mathcal{X}}_{4,1,3}$ is therefore a sphere $S^2$ with eight points missing like holes. By imagining stretching one hole out to infinity, flattening the sphere to a disk, and then contracting the space between the remaining seven holes, one can demonstrate that $\overline{\mathcal{X}}_{4,1,3}$ is topologically equivalent to the wedge product of seven circles, i.e.\ seven loops that share a single point. %Each loop defines an independent generator of $\pi_1(\overline{\mathcal{X}}_{4,1,3})=\pi_1(\mathcal{X}_{4,1,3})=PT_4$.

\section{Fundamental groups}\label{sect:fund}

The space $\mathcal{X}_{N,d,k}$ is a manifold for any $N$, $d$ and $k$, and the usual definition of the fundamental group can be applied to describe its topology~\cite{hatcher_algebraic_2001}. As stated before, only $\mathcal{X}_{N,2,2}$ and $\mathcal{X}_{N,1,3}$ have non-trivial fundamental groups $PB_N = \pi_1(\mathcal{X}_{N,2,2})$ and $PT_N = \pi_1(\mathcal{X}_{N,1,3})$. The lowest possible manifestations of the two groups are isomorphic $PB_2 = PT_3 =\Z$, but they are different groups after that. The next traid group $PT_4 = \pi_1(\mathcal{X}_{4,1,3})= \pi_1(\overline{\mathcal{X}}_{4,1,3}) =F_7$, the free group with seven generators. The number of generators of $PT_N$ is at least $2^{N-3} (N^2 - 5N +8) -1$, the Betti number for $\mathcal{X}_{N,1,3}$~\cite{bjorner_homology_1995,barcelo_discrete_2008}. For $N \geq  6$, the existence of disjoint triplets of particles imply $PT_N$ is no longer a free group.

Similar to the relation between the pure braid group and braid group, the pure traid group is the subgroup of the traid group that contains by all elements that do not permute the particles. The traid group $T_N$ has a much simpler structure than $PT_N$ and only $N-1$ generators for all $N$. Ideally we would like to define $T_N$ as the fundamental group of $\mathcal{X}_{N,1,3}/S_N$ and use it to analyze the subgroup $PT_N$. However, a complicating factor is that the quotient space $\mathcal{X}_{N,d,k}/S_N$ is only a manifold when $k=2$. When $k=2$, the action of any non-trivial element of $S_N$ takes some open neighborhood of any point entirely off of itself. 
Therefore there is no problem defining the braid group $B_N$ as the fundamental group of $\mathcal{X}_{N,2,2}/S_N$. However, when $k>2$ the configuration space contains points with a repeated coordinate. For these points, there are non-trivial elements of the symmetric group which fix the point and act on open neighborhoods by a reflective symmetry. Therefore $\mathcal{X}_{N,d,k}/S_N$ is not a manifold and taking the quotient na\"{i}vely loses important topological information. For example, $\overline{\mathcal{X}}_{3,1,3}/S_3$ is a closed line segment and $\overline{\mathcal{X}}_{4,1,3}/S_4$ is a triangle missing two corners. To preserve the topological information, we must regard the quotient as an {\em orbifold} and define a generalization of the fundamental group called the \emph{orbifold fundamental group}~\cite{thurston_geometry_2002}. For details, see the Appendix.

With this generalized definition, the orbifold fundamental group $T_N = \pi_1(\mathcal{X}_{N,1,3}/S_N)$ is  the semidirect product $T_N = PT_N \ltimes S_N$ of the pure traid group $PT_N$ and the symmetric group $S_N$. The corresponding short exact sequence is $1 \to PT_N \into T_N \twoheadrightarrow S_N \to 1$.

\section{Presentations}\label{sect:pres}

\begin{figure}
\centering
 \includegraphics{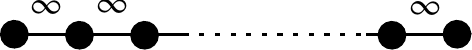}
 \caption{The Coxeter-Dynkin diagram for $T_N$ has $N-1$ nodes arranged linearly with all labels equal to $\infty$. Any of the finite-labeled linear Coxeter groups on $N-1$ generators can be considered a quotient of $T_N$. When the finite labels are multiples of $3$, the quotient is a normal subgroup contained within $PT_N$. } \label{fig:coxeter}
\end{figure}

The group $T_N$ is generated by elementary moves which pass pairs of adjacent particles through each other; these correspond to paths in the configuration space $\mathcal{X}_{N,1,3}$ which return the individual particles to their original positions as a set and which cross $V_{N,1,2}$ at a single generic point. In contrast with the braid group, there is no over/under crossing information. Since two-particle coincidences are allowed, each of the elementary swaps are square trivial and are thus their own inverse; see Fig.~\ref{fig:braidtraidcontrast}. Further, the braid group allows the third Reidemeister move (a Yang-Baxter relation from knot theory) as shown in Fig.~\ref{fig:braidtraidcontrast}, which would introduce a triple point in our context. It follows that the traid group $T_N$ has a presentation with generators, $t_1,\ldots, t_{N-1}$, and relations:
\begin{eqnarray}\label{eq:tpres}
   t_i^2 & = 1 \ &\forall i \nonumber \\
   t_it_j & = t_jt_i \ &\forall |i-j|>1 .
\end{eqnarray} 
Each of the $N-1$ generators correspond to trades of the $i^{th}$ and $(i+1)^{th}$ elements. As in the braid group, all sufficiently distant generators commute. The first defining relation of the traid group $t_i^2=1$ for all $i$ means that the traid group is generated by reflections. Combined with the second defining relation of the traid group, which can be rewritten to $(t_it_j)^2=1$ for all $|i-j|>1$, mean that $T_N$ is a \emph{linear} Coxeter group, i.e.\ it does not branch or loop. See Fig.~\ref{fig:coxeter}.

It is interesting to note the relationships between the presentations of the traid, symmetric, and braid groups on $N$ strands. We have two slightly different presentations for the symmetric group. The first has generators $s_1, \ldots, s_{N-1}$ and relations
\begin{align*}
 s_i^2 & = 1 & \forall i \\
 s_i s_j & = s_j s_i & \forall |i-j|>1 \\
 (s_{i+1} s_i)^3 & = 1 & \forall i< N-1
\end{align*}
The traid group is then the group which is obtained by erasing the triple point relation $(s_{i+1}s_i)^3=1$. 
The other symmetric group presentation rewrites the triple point relation into the Yang-Baxter relation:
\begin{displaymath}
 \Big[ (s_{i+1} s_i)^3 = 1 \Big] \to \Big[ s_is_{i+1}s_i = s_{i+1}s_is_{i+1} \Big]
\end{displaymath}
The braid group has a presentation with generators $b_1,\ldots, b_{N-1}$ and relations
\begin{align*}
 b_i b_j & = b_j b_i & \forall |i-j|>1 \\
 b_i b_{i+1} b_i & = b_{i+1} b_i b_{i+1} & \forall i< N-1
\end{align*}
which omits only the $s_i^2=1$ relation from the symmetric group. As with the traid group, we get a homomorphism $B_N \twoheadrightarrow S_N$ induced by $b_i \mapsto s_i$ whose kernel is the pure braid group $PB_N$. The preservation of the Yang-Baxter relation is important and is related to the realization of $B_N$ as the mapping class group of an $N$-fold marked disc. The traid groups have no such realization as an automorphism of the underlying space.

\begin{figure}
\centering
 \adjustbox{trim={.12\width} {.12\height} {0.12\width} {.10\height},clip}{\includegraphics[width= 1.2\columnwidth]{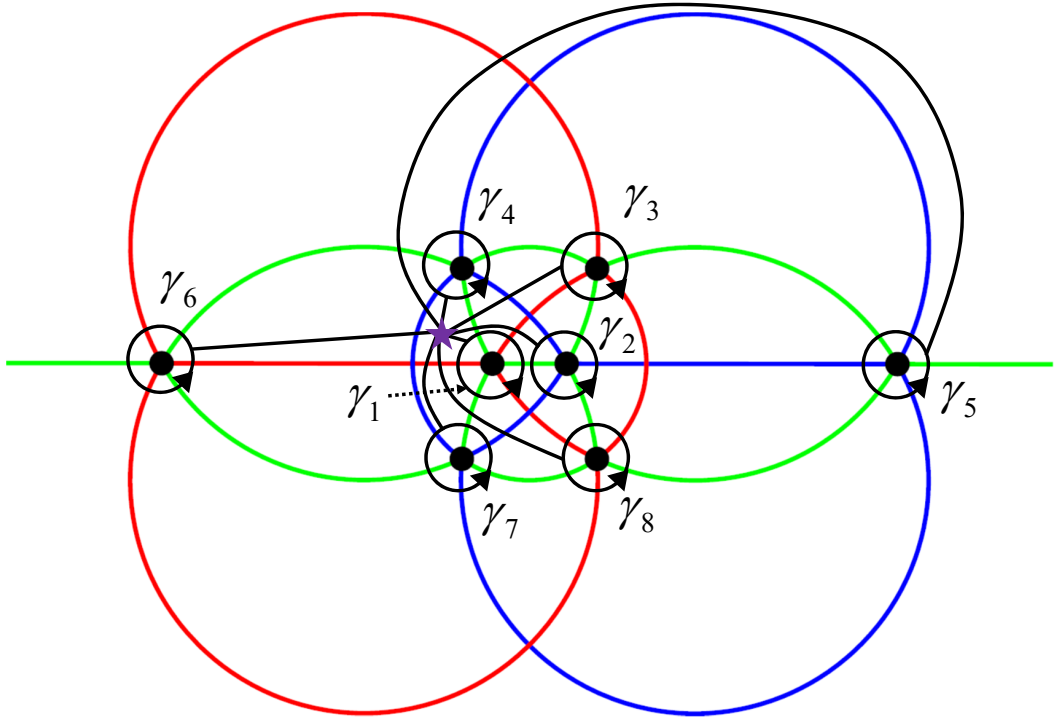}}
 \caption{This figure depicts eight loops $\gamma_1$ through $\gamma_8$ that provide generators for the presentation of $PT_4$ described in Eqs.~\ref{eq:pt4}. The purple star is an arbitrarily chosen base point in $\overline{\mathcal{X}_{4,1,3}}$ (see Fig.~\ref{fig:reducedX413}) where all loops start and end and the direction of all loops travel out from the base point, then go counterclockwise around the triple coincidence point, and then return back along the same path. The colored circles correspond to the two-particle coincidences, but here they are colored by the elements of the traid group. The red arcs correspond to $t_1$ (i.e.\ exchanging the first and second particle, no matter which particle they are), the green are $t_2$, and the blue are $t_3$. For example, loop $\gamma_1$ crosses red, green, red, green, red and then green, and so $\gamma_1 = t_2 t_1 t_2 t_1 t_2 t_1 =(t_2 t_1)^3$. Loop $\gamma_2$ crosses green, red, green, blue, green, blue, green, blue, red, then finally green again and so $\gamma_2 = t_2t_1(t_3 t_2)^3 t_1t_2$.} \label{fig:pt4}
\end{figure}

As the pure traid group is precisely the kernel of the homomorphism $T_N\twoheadrightarrow S_N$ given by $t_i\to s_i$ on the generators, the pure traid group $PT_N$ can be seen to be normally generated in $T_N$ by the elements of the form $(t_{i+1}t_i)^3$. For example, $PT_3$ is a copy of $\Z$ generated by $(t_2t_1)^3$. The next pure traid group $PT_4$ is somewhat more complicated to express in terms of traid group generators. One possible presentation of $PT_4$ has eight generators $\gamma_1,\ldots,\gamma_8$ and one relation
\begin{equation*}
 \gamma_8 \gamma_7 \gamma_6 \gamma_5 \gamma_4 \gamma_3 \gamma_2 \gamma_1 = 1.  
\end{equation*}
 The eight generators of $PT_4$ can be constructed as products of the three generators of $T_4$:
\begin{equation}
\begin{aligned}
 \gamma_1 & = & (t_2t_1)^3 &&  \gamma_5 & = & t_3t_2t_1(t_2t_3)^3t_1t_2t_3\\
 \gamma_2 & = & t_2t_1(t_3t_2)^3t_1t_2 &&  \gamma_6 & = & t_3(t_1t_2)^3t_3\\
 \gamma_3 & = & t_2t_3(t_2t_1)^3t_3t_2 &&  \gamma_7 & = & t_1(t_2t_3)^3t_1\\
 \gamma_4 & = & (t_3t_2)^3 && \gamma_8 & = & t_1t_2t_3(t_1t_2)^3t_3t_2t_1
\end{aligned}\label{eq:pt4}
\end{equation}
The order of the traid group generators $t_i$ can be inferred from the pattern of paths in Fig.~\ref{fig:pt4}.

\section{Representations}\label{sect:repres}

Because $t_i^2=1$ for each of the generators, the abelian representations of $T_N$ are easily classified. Any representation $\rho:T_N\to U(1)$ must have $\rho(t_i)=\pm 1$ for all $i$, and then all other constraints of the presentation (\ref{eq:tpres}) are satisfied. Consequently, there are $2^{N-1}$ abelian representations of $T_N$ corresponding to all binary choices of signs for each generator\footnote{In this paper, we have not considered projective representations of the traid group, only unitary representations.}. The simplest two are when $\rho(t_i)=+ 1$ for all $i$ (this is equivalent to the bosonic representation of $S_N$) and when $\rho(t_i)=- 1$ for all $i$ (this is equivalent to the fermionic representation of $S_N$). Novel abelian representations of $T_N$ that cannot be factored through $S_N$ representations occur when there are mixed signs. 

\begin{figure}
\centering
\includegraphics[width= .9\columnwidth]{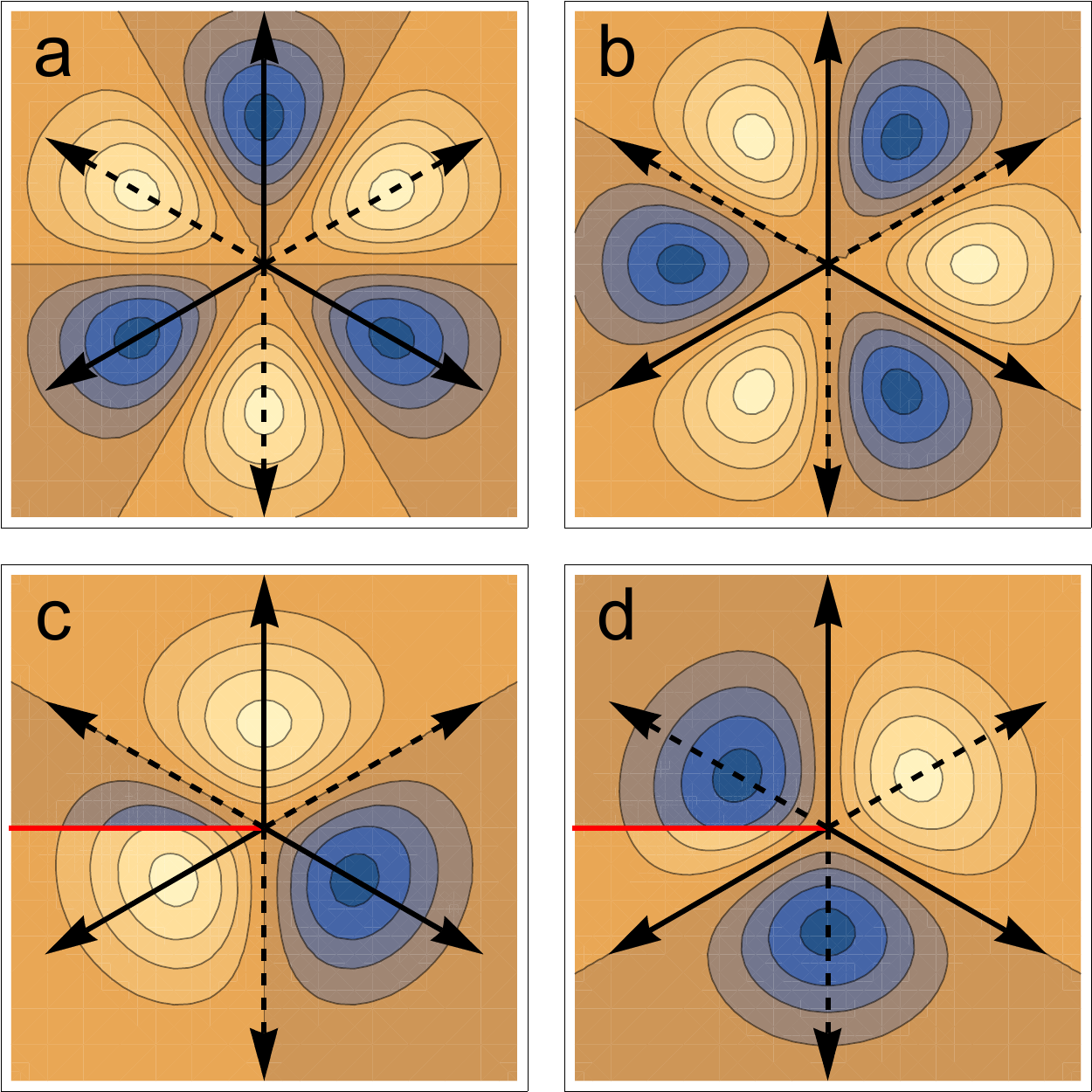}
 \caption{Subfigures a and b are contour plots of the relative wave function of the lowest energy eigenstates of $H_\mathrm{rel}$ (\ref{eq:2dham}) with $\lambda = 3$ for bosons $\rho(t_1)=\rho(t_2) = +1$ and fermions $\rho(t_1)=\rho(t_2) = -1$, respectively. Subfigures c and d depict the relative wave function of the lowest energy eigenstates of $H_\mathrm{rel}$ with $\lambda = 3/2$ and anyonic symmetry in the abelian representations $\rho(t_1)=-\rho(t_2) = +1$ and $\rho(t_1)=-\rho(t_2) = -1$, respectively. The solid arrows correspond to the boundaries defined by the $t_1$ generators and the dashed arrows to boundaries the $t_2$ generators. The red horizontal line in subfigures c and d is the branch cut arising from the double-covering created by fractional angular momentum; the wave function changes sign when crossing this branch.} \label{fig:states}
\end{figure}

For an example where there are wave functions that transform like these mixed representations, consider the simplest case of three identical particles in a one-dimensional harmonic trap with zero-range hard-core three-body interactions. The Hamiltonian for this system can be expressed in scaled particle coordinates as
\begin{equation}
H = \frac{\hbar \omega}{2} \sum_{i=1}^3 \left( -\frac{\partial^2}{\partial x_i^2} + x_i^2 \right) + g \delta(x_1-x_2)\delta(x_2-x_3)
\end{equation}
in the limit $g \to \infty$. Transforming to Jacobi polar relative coordinates
%\begin{eqnarray}
%q_1 &=& \frac{1}{\sqrt{2}} (x_1 - x_2)\nonumber\\
%q_2 &=& \frac{1}{\sqrt{2}} (x_1+x_2 -2 x_3),
%\end{eqnarray}
$\rho$ and $\varphi$~\cite{harshman_symmetries_2012}
\begin{eqnarray}
\rho^2 &=& \frac{2}{3} \left(x_1^2 + x_2^2 + x_3^2 -x_1x_2 -x_2x_3 - x_3x_1\right)\nonumber\\
\tan \varphi &=& \frac{\sqrt{3} (x_1 - x_2)}{x_1+x_2 -2 x_3},
\end{eqnarray}
the relative Hamiltonian becomes
\begin{equation}\label{eq:2dham}
H_\mathrm{rel} = \frac{\hbar \omega}{2} \left[ -\frac{1}{\rho}\frac{\partial}{\partial \rho}\left(\rho \frac{\partial}{\partial \rho}\right) - \frac{1}{\rho^2} \frac{\partial^2}{\partial \varphi^2}+ \rho^2 \right] + g \delta^{(2)}(\rho)
\end{equation}
(again in the limit $g\to \infty$). Note that this has the same functional form as the relative Hamiltonian for two-particles with zero-range hard-core two-body interactions in a two-dimensional harmonic trap, a perennial test-bed for studying fractional exchange statistics, c.f.~\cite{leinaas_theory_1977,wilczek_quantum_1982,khare_fractional_1997}. The energy spectrum for the relative Hamiltonian (\ref{eq:2dham}) is $E_\mathrm{rel} = \hbar\omega (2 \nu + \lambda + 1)$, where $\nu$ is a non-negative integer counting the radial nodes and $\lambda$ is the relative `angular momentum'. The spectrum of allowed values of $\lambda$ depends on the boundary conditions implied by the particle statistics. For bosonic $\rho(t_1)=\rho(t_2)=+ 1$ and fermionic $\rho(t_1)=\rho(t_2)=- 1$ representations, the magnitude of the three-body relative `angular momentum' $\lambda$ is restricted to values that are positive integer multiples of $3$~\cite{harshman_symmetries_2012}. However, states with relative angular momenta $\lambda = 3/2$, $9/2$, etc.\ satisfy the exchange statistics governed by the mixed abelian representations of $T_3$ where $\rho(t_1)=-\rho(t_2)$. These wave functions are double-valued on $\mathcal{X}_{3,1,3}$, but they give single-valued, anyonic wave functions on the orbifold $\mathcal{X}_{3,1,3}/S_3$ (see Fig.~\ref{fig:states}). The states have lower energy than the lowest energy fermionic and bosonic states. Extending these preliminary results and comparing to the case of braid anyons and gauge transformations is ongoing work.

Any representation of $T_N$ gives a representation of $PT_N$ by restriction. Mixed abelian representations of $T_N$ restrict to non-trivial representations of $PT_N$ in which a cyclic exchange of distinguishable particles leads to a sign change of the wave functions. Additionally, there are representations of $PT_N$ which do not extend to representations of $T_N$. For example, $PT_3\sim \Z$ and so has infinitely many abelian representations $\rho_\theta(n)=e^{in\theta}$ parametrized by $\theta\in[0,2\pi)$, similar to the fractional exchange statistics. Further, $PT_4$ has an abelian representation for any assignment $\rho(\gamma_k)=e^{i\theta_k}$ with $\sum_{k=1}^8 \theta_k = 0 \mod 2\pi$. In other words, unlike abelian representations of $B_N$, not all generators must have the same phase. In general, abelian representations of $PT_N$ must factor through the abelianization and are parametrized by the first cohomology of the configuration space with coefficients in $U(1)$, $H^1(\mathcal{X}_{N,1,3};U(1))$. The integer versions of these groups have been calculated in \cite{bjorner_homology_1995}. In particular, as the first homology is non-trivial for $N\geq 3$, $PT_N$ always has many abelian representations.

Non-abelian representations of $T_N$ and $PT_N$ are a seemingly rich topic. As shown in Fig.~\ref{fig:coxeter}, the group $T_N \sim [\infty^{N-1}]$ may be considered a kind of `universal object' in the category of Coxeter groups with linear Coxeter-Dynkin diagrams. In other words, for every irreducible representation of any linear Coxter group with $N-1$ generators, one can induce an irreducible representation of $T_N$. As an example, consider the case when $N=4$ and define $K^{[m,n]}$ as the normal subgroup of $T_4$ generated by $(t_2t_1)^m$ and $(t_3t_2)^n$ and their conjugates by elements of $T_4$. Note that this means that the normal subgroup includes the terms generated by inequivalent conjugates like those found in (\ref{eq:pt4}). Then the quotient group $T_4/K^{[m,n]}$ is the linear Coxeter group $[m,n]$. Examples include
\begin{itemize}
\item the finite Coxeter groups $[3,3]\equiv A_3 \sim S_4$, $[4,3] \equiv C_3$, and $[5,3] \equiv H_3$,
\item the affine Coxeter groups $[4,4]\equiv \tilde{C}_2$ and $[6,3]\equiv \tilde{G}_2$, and
\item the hyperbolic Coxeter groups $[p,q]$ with $2(p+q)<pq$.
\end{itemize}
Using the homomorphism $T_4 \rightarrow [m,n]$, one can construct a representation of $T_4$ from any irreducible representation of $[m,n]$ by pullback, including multi-dimensional, non-abelian representations. In particular, note that the special case $K^{[3,3]} = PT_4$ leads to the symmetric group $[3,3]\equiv A_3 \sim S_4$. This method extends to any $N$, but at this point, it is not known whether the irreducible representations constructed this way exhaust the irreducible representations of $T_N$.

As for non-abelian representations of $PT_N$ for $N>3$, we similarly expect there to be many. The groups $PT_4$ and $PT_5$ are free and so homomorphisms to any group exist and the generators of the $PT_N$'s may be sent to arbitrary values. When $N\geq 6$, there are additional commutation relations among the generators which need to be satisfied, but we have not yet determined the full set of relations.

\section{Summary and outlook}\label{sect:summ}

In summary, the anyonic physics that derives from the not-simply-connected configuration space of $N$ particles with three-body hard-core interactions in one-dimension has some similarities but also intriguing differences from the more familiar hard-core two-body interactions in two dimensions. Similarities include: 1) the braid group $B_N$ and the traid group $T_N$ are both generalizations of the symmetric group $S_N$ with one defining relation removed; 2) both $B_N$ and $T_N$ have `pure' subgroups describing distinguishable particles; 3) representations of both groups give abelian and non-abelian generalized exchange statistics; and 4) unlike fermionic and bosonic wave functions, anyonic wave functions cannot be built from the tensor product of one-particle states.

Differences between braid anyons and traid anyons include: 1) in the traid group, the Yang-Baxter relation is broken instead of the square-trivial relation as in the braid group; 2) the traid group does not have an interpretation in terms of diffeomorphisms of the underlying space; and 3) the traid group derives from the orbifold fundamental group of the identical particle configuration space instead of the more familiar notion of the fundamental group. 

Two extensions of this work immediately suggest themselves. First, the representation theory of the traid groups is far from developed, and that will be necessary before model-building and analysis can elucidate the differences from braid anyons more clearly and before implementations and observables can be suggested for experiments with ultracold atoms. For example, another difference between braid anyons and traid anyons worth more exploration is what happens when parity reflections are included. Traid anyons in a parity symmetric trap should either be parity-symmetric or come in doublets that mix under parity. In contrast, braid anyons do not respect parity~\cite{khare_fractional_1997}. The mixing of traid group representations under supersymmetry also appears to be an interesting question. Second, for braid anyons, the generalized exchange statistics can be incorporated into an interaction derived from a gauge field~\cite{wilczek_quantum_1982,wilczek_fractional_1990,forte_quantum_1992,khare_fractional_1997}. The preliminary results for traid anyons in a one-dimensional harmonic trap mentioned above suggest that a similar transmutation of statistics into non-local few-body interactions is possible.

\emph{Acknowledgments.} NLH would like to thank D.\ Blume, S.\ Chandrasekharan, A.\ Eckhardt, M.A.\ Garci\'{a}-March, J.M.\ Midtgaard, M.\ Olshanii, and N.T.~Zinner for helpful discussions.

\section*{Appendix: Orbifolds and Orbifold Fundamental Groups}

In order to be self-contained, we begin with a review of some basic constructions of algebraic topology with a view toward explaining their less frequently encountered orbifold generalizations.

\subsection{Fundamental groups}

Suppose that $X$ is a locally path connected space and that we have selected a {\em base point} $x \in X$. The {\em fundamental group}, $\pi_1(X,x)$, is the group of homotopy classes of paths $\gamma:[0,1]\to X$ which begin and end on $x$, i.e.\  $\gamma(0)=\gamma(1)=x$. Two such paths $\gamma_0,\gamma_1$ are considered homotopic when they can be continuously deformed to one another. More precisely, there should exist a map $H:[0,1]\times [0,1] \to X$ such that $H(0,t)=\gamma_0(t)$, $H(1,t)=\gamma_1(t)$, and $H(s,0)=H(s,1)=x$. The space of based loops naturally splits into connected components by homotopy equivalence.

Multiplication in $\pi_1(X,x)$ is the concatenation operation. That is, $\gamma_1\gamma_0$ is the path which runs through $\gamma_0$ on $[0,1/2]$ and through $\gamma_1$ on $[1/2,1]$, each twice as fast as originally. Associativity does not hold `on the nose', but up to parametrization of $[0,1]$ realizable as a homotopy. Similarly, $\gamma^{-1}$ can be identified with a copy of $\gamma$ given the reversed parametrization.

Although the base point $x$ is necessary for the definition of $\pi_1(X,x)$, the isomorphism type of the fundamental group is independent of $x$ whenever $X$ is path connected. In particular, if we are given two base points $x_0,x_1$ and a path $\gamma_*$ from $x_0$ to $x_1$, we can explicitly describe the isomorphism of $\pi_1(X,x_0) \to \pi_1(X,x_1)$ as being induced by the map on paths $\gamma \to \gamma_* \gamma \gamma_*^{-1}$. (If $x_0=x_1$, this gives an action of $\pi_1(X,x_0)$ on itself by conjugation.) When the base point is not of fundamental importance, we will sometimes omit it and simply write $\pi_1(X)$.

The fundamental group is covariantly functorial in the sense that, if we are given a map of pointed spaces $f:(X,x) \to (Y,y)$, we get a homomorphism $f_*:\pi_1(X,x) \to \pi_1(Y,y)$. Further, homotopy equivalent spaces have isomorphic groups.

There are several techniques for computing the fundamental group of a space. Some can be computed by considering group actions and covering spaces (see below). In particular, we can determine $\pi_1(S^1)\cong \Z$ this way. Also, we can form the {\em wedge sum} of pointed spaces $(X,x)$ and $(Y,y)$ in which we take a copy of $X$ and a copy of $Y$ and identify their base points $x$ and $y$. The resulting pointed space $(X \vee Y,x=y)$ has fundamental group $\pi_1(X\vee Y, x=y)$ equal to the free product $\pi_1(X,x)*\pi_1(Y,y)$. This is a specific case of the more general van-Kampen Theorem~\cite{hatcher_algebraic_2001}.

%These facts allow us to compute
%\begin{itemize}
%\item $\pi_1({\cal X}_{3,1,3})=\Z$ since ${\cal X}_{3,1,3}$ is homotopy equivalent to $S^1$, and
%\item $\pi_1({\cal X}_{4,1,3})=F_7$ (the free group on seven letters) since ${\cal X}_{4,1,3}$, being homotopy equivalent to the 2-sphere minus 8 points, is homotopy equivalent to the wedge sum of seven circles.
%\end{itemize}

Now consider the set $\pi_1(X,x_0,x_1)$ of homotopy types of paths from $x_0\in X$ to $x_1 \in X$. Equivalently, $\pi_1(X,x_0,x_1)$ is the set of connected components ($\pi_0$) of the same space of paths. This space of paths shows up as the domain of integration for the path integral, so it is of some importance. In particular, when $\pi_1(X,x_0,x_1)$ is non-trivial, the domain of the path integral splits into several connected components. So we have some interest in determining its structure.

Perhaps unsurprisingly, as a set $\pi_1(X,x_0,x_1)$ is in bijective correspondence with $\pi_1(X,x)$. More strongly, $\pi_1(X,x_0,x_1)$ is affine equivalent to $\pi_1(X)$ in the sense that if we choose any $\gamma_*\in \pi_1(X,x_0,x_1)$, we get a map $\pi_1(X,x_0,x_1)\to \pi_1(X,x_0)$ by $\gamma \to \gamma_*^{-1} \gamma $.

\subsection{Covering groups}

Suppose that we are given two topological spaces, $X$ and $Y$, and a map between them, $p:X \to Y$. We say that $X$ {\em covers} $Y$ via $p$ if every point $y\in Y$ has some neighborhood $U_y$ for which $p^{-1}(U_y)$ is the disjoint union of sets, each of which is mapped homeomorphically onto $U_y$ via $p$. (In particular, if either of $X$ or $Y$ are manifolds, then so is the other.)

A classic set of examples are the map $i\R \to U(1)=S^1$ given by the exponential and the $n^{th}$ power maps $U(1)\to U(1)$ given by $z \mapsto z^n$.

A common source of covering spaces is group actions. That is, suppose that some group $G$ acts on $X$ by homeomorphisms. Then the quotient map $p: X \to X/G$ defines a covering space exactly when, for every point $x \in X$, the action of $g\in G$ either acts as the identity on a neighborhood of $x$ or takes some neighborhood of $x$ entirely off of itself, i.e.\ a properly discontinuous group action. From the standpoint of the covering, $G$ acts by {\em deck transformations} of $p:X \to Y$. i.e. homeomorphisms $D: X \to X$ for which $p\circ D = p$.

An example of such a group action comes from the $S_N$ action on the configuration spaces ${\cal X}_{N,d,2}$. Since the locus $V_{N,d,2}$ of two coincidences has been removed, each of the coordinates $x_i \in \R^d$ in $x=(x_1,\ldots, x_{N})$ occurs at most once. Thus every non-identity element of $S_N$ will take some neighborhood of $x$ completely off of itself. We return to this example in the section on orbifolds, below.

The set of deck transformations form a group in any case, but when the (injective) image $p_*\pi_1(X)$ is a normal subgroup of $\pi_1(Y)$, the quotient $\pi_1(Y)/p_*\pi_1(X)$ is isomorphic to the group of deck transformations. Such coverings are called normal (or regular) coverings.

Thus, for normal coverings $X \to Y$, we have an exact sequence of groups
\begin{displaymath}
 0 \to \pi_1(X) \to \pi_1(Y) \to D \to 0
\end{displaymath}
where $D$ is the group of deck transformations. Equivalently, $\pi_1(Y)$ is a semidirect product of $\pi_1(X)$ and $D$.

To provide eventual contrast between covering spaces and orbifolds, we explore the map $\pi_1(Y) \to D$. 
Suppose that we choose $y \in Y$ and let $p^{-1}(y) = \{x_1, \ldots, x_N\}$. As every point $y' \in Y$ has a neighborhood $U_{y'}$ for which $p^{-1}(U_{y'})$ is simply a collection of disjoint copies of $U_{y'}$, we have what is called the \emph{unique lifting property}. That is, given any contractable set $(C,c)$ and $f:(C,c) \to (Y,y)$, there is a unique lift $\tilde{f}:(C,c) \to (X,x_i)$ for each $x_i \in p^{-1}(y)$. More concretely, every loop in $\pi_1(Y,y)$ lifts uniquely to a path in $X$ between points of $p^{-1}(y)$, once you say where it begins. Further, every homotopy of paths lifts uniquely under the same conditions.

This determines a group action of $\pi_1(Y,y)$ on the fiber $p^{-1}(y)$ by permutations. In other words, given $\gamma \in \pi_1(Y,y)$ the action of the permutation $s$ on $x_i$ is given by lifting $\gamma$ with starting point $x_i$ and observing where the other end point is. The loops of $p_*\pi_1(X)$ lift to closed loops and induce trivial permutations of $p^{-1}(y)$.

The deck transformation group acts similarly on $p^{-1}(y)$ and the map $\pi_1(Y) \to D$ is the one which identifies the element of $\pi_1(Y)$ with its corresponding permutation.

\subsection{Orbifolds}

Before defining orbifolds, let us revisit hard-core configuration spaces in the case $k=2$. For the space ${\cal X}_{N,d,2}$, the $S_N$ group action is properly discontinuous and so ${\cal X}_{N,d,2} \to {\cal X}_{N,d,2}/S_N$ is a covering space. As the locus $V_{N,d,2}$ is a co-dimension $d$ set, its removal affects $\pi_0$ when $d=1$, $\pi_1$ when $d=2$, and higher homotopy groups when $d>2$. When $d=2$, the map ${\cal X}_{N,2,2} \to {\cal X}_{N,2,2}/S_N$ is that which passes from the configuration space of the pure braid group to that of the braid group. The induced map on $\pi_1$ is the inclusion $PB_N\into B_N$.

When $d=1$, we see that ${\cal X}_{N,1,2}$ is a collection of $N!$ disconnected contractable sets each labeled with a permutation of the set $\{1,\ldots,N\}$. The connected components of $\overline{{\cal X}}_{N,1,2}$ are open $N-2$ simplices. The action of $S_N$ is simply the permutation action on the labels, so the covering ${\cal X}_{N,1,2} \to {\cal X}_{N,1,2}/S_N$ is the trivial covering of the base by a number of disjoint copies of itself.

\begin{figure}
  \includegraphics[scale=0.4]{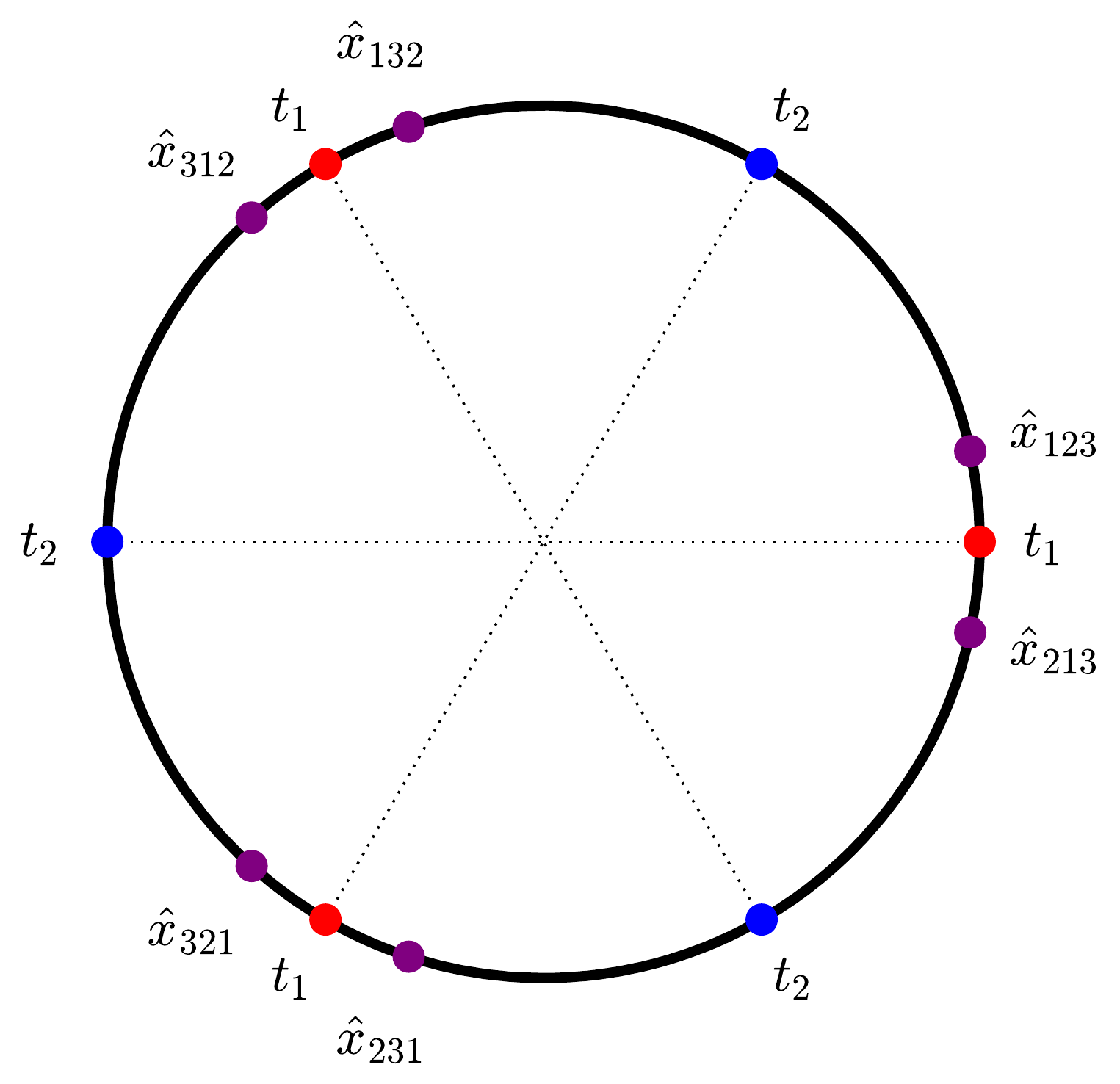}
  \caption{The reduced configuration space $\overline{\mathcal{X}}_{3,1,3}$ with 2-coincidence locus in $\overline{\mathcal{V}}_{3,1,2}$ colored in red and blue. Labels on the 2-coincidences correspond to generators of the traid group $T_3$. In purple, $3!$ base points which are identified under the $S_3$ action. } \label{fig:X313}
\end{figure}

\begin{figure}
  \includegraphics[scale=0.4]{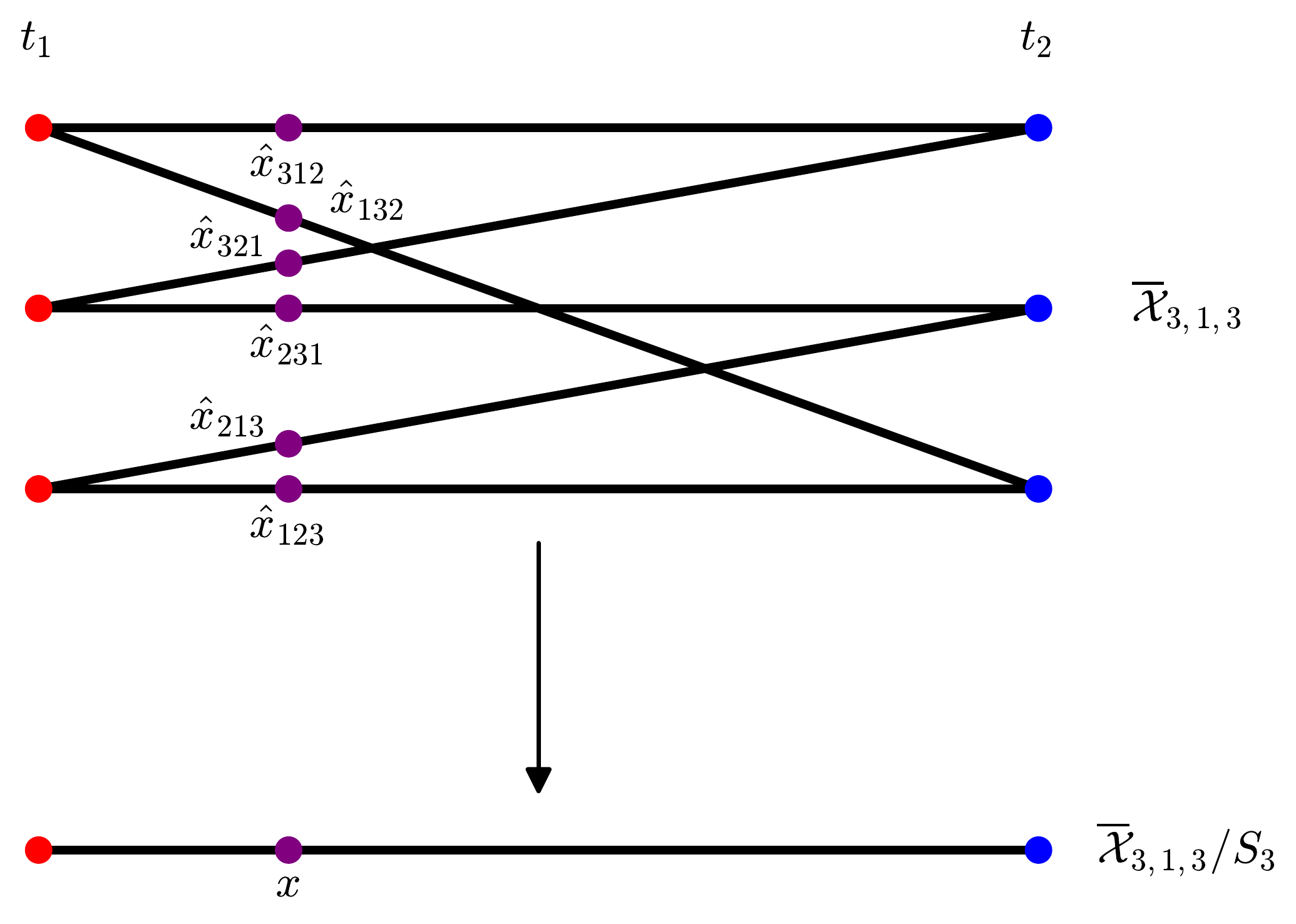}
  \caption{The quotient map $\overline{\mathcal{X}}_{3,1,3}\to \overline{\mathcal{X}}_{3,1,3}/S_3$ with 2-coincidence locus in red and blue. In purple, $3!$ base points which are identified under the $S_3$ action. } \label{fig:quotientX313}
\end{figure}

Now let us consider the $k=3$ case. Since $V_{N,d,k'} \subset V_{N,d,k}$ for $k'>k$, ${\cal X}_{N,d,k} \subset {\cal X}_{N,d,k'}$. Thus ${\cal X}_{N,d,2} \subset {\cal X}_{N,d,3}$ and we should consider ${\cal X}_{N,d,3}$ as being formed by taking ${\cal X}_{N,d,2}$ and adding in the co-dimension $2d$ two-but-not-three coincidence locus $V_{N,d,2} - V_{N,d,3}$.

The key observation is that the $S_N$ action, and the quotient by it, no longer correspond to a covering space. The points of $V_{N,d,2}- V_{N,d,3}$ have at least one repeated coordinate and elements of the symmetric group which are transpositions of those two coordinates fix those points {\em but do not act as the identity in any neighborhood of those points.}

In fact, something very bad happens to the na\"{i}ve topology. As an instructive example, consider $\overline{\mathcal{X}}_{3,1,3}$ as shown in Fig.~\ref{fig:X313}. A loop going around the circle once describes the generator of $PT_3\cong \Z$ and corresponds to the choreography of 3 particles seen in Fig.~2 of the main text. The quotient map is diagrammed in Fig.~\ref{fig:quotientX313}. Of particular note is that, as a bare topological space, the quotient is an interval and the fundamental group is trivial.

To retain the topological information we found in $\mathcal{X}_{N,1,3}$, we will need to consider the quotient space $\mathcal{X}_{N,1,3}/S_N$ to be an {\em orbifold}. Orbifolds are generalizations of manifolds where every point has a neighborhood which is modeled on $\R^n$ but these charts may come equipped with a possibly non-trivial group of enforced symmetries.

For $\mathcal{X}_{N,1,3}$, generic points (corresponding to $\mathcal{X}_{N,1,2}$) have usual manifold charts with only the identity as a symmetry. The points of $\mathcal{X}_{N,1,3}\cap V_{N,1,2}$, however, come with one or more additional symmetries: a reflective symmetry for each pair of coincident particles. (These are commonly known as orbifold points as opposed to the trivial-symmetry points which are called manifold points.) As each pair of coincident particles must be disjoint (it would be in $V_{N,1,3}$ otherwise) these symmetries commute. This type of intersection can be seen in Fig.~3 of main text, at the two-fold intersections of the circles of $\overline{V}_{4,1,2}$.

Generally, we hope to get a orbifold version of the covering space construction and its algebraic implications. Unfortunately, we cannot hope to simply copy the covering space material above {\em mutatis mutandis}. The issue is that, due to the presence of orbifold points, we cannot hope to get the unique lifting property. As an example, consider the quotient map in Fig.~\ref{fig:quotientX313}. The quotient space has exactly two orbifold points, in red and blue. Suppose that we were to take a path which starts at $x$, goes to the red orbifold point and then returns to $x$. How are we to lift it to a path in $\overline{\mathcal{X}}_{3,1,3}$ starting at $\hat{x}_{123}$? There are now two options:
\begin{itemize}
\item we can lift it to a path which starts at $\hat{x}_{123}$, goes through the red point, and proceeds to $\hat{x}_{213}$, or
\item we can lift it to a path which starts at $\hat{x}_{123}$, touches the red point, and returns to $\hat{x}_{123}$.
\end{itemize}

In order to properly generalize the notion of $\pi_1$ to orbifolds, we would need to use a groupoid, which is a type of group-like object where multiplication is only partially defined. However, for the type of orbifold given by $\mathcal{X}_{N,1,3}/S_N$, i.e.\ the global quotient of a manifold by a finite group acting by diffeomorphisms faithfully on an open dense set, we can assign an group in the following way: Let $x$ be a point of $\mathcal{X}_{N,1,3}$ at which $S_N$ acts faithfully (a point of $\mathcal{X}_{N,1,2}$), so that if $p:\mathcal{X}_{N,1,3} \to \mathcal{X}_{N,1,3}/S_N$ is the quotient projection, then $p^{-1}(x)$ is a full set of $N!$ points, i.e.\ the stabilizer of $x$ is trivial. We choose an arbitrary point in $p^{-1}(x)$ and label it by the trivial permutation $\hat{x}_{1\cdots N}$; the other points $\hat{x}_s=s\hat{x}_{1 \cdots N}$ are then identified with their corresponding permutations $s\in S_N$. Let the orbifold fundamental group $\pi_1(\mathcal{X}_{N,1,3}/S_N,x)$ be defined as the set of maps $\gamma:[0,1]\to \mathcal{X}_{N,1,3}$ with $\gamma(0)=\hat{x}_{1 \cdots N}$ and $\gamma(1)\in p^{-1}(x)$, modulo boundary-relative homotopy.

The multiplication on this group is defined as follows: Suppose that $\gamma, \gamma'\in \pi_1(\mathcal{X}_{N,1,3},x)$ so that $\gamma,\gamma'$ begin at $\hat{x}_{1 \cdots N}$ but end at $\hat{x}_s, \hat{x}_{s'} \in p^{-1}(x)$, respectively.  
Then $\gamma' \gamma$ is defined to be the concatenation of $s\gamma'$ and $\gamma$, a path which begins at $\hat{x}_{1 \cdots N}$ and ends at $\hat{x}_{s's}$.

Similar to the theory of covering spaces, we can naturally regard the orbifold fundamental group 
$T_N = \pi_1(\mathcal{X}_{N,1,3}/S_N,x)$ as a semidirect product of the fundamental group 
$PT_N=\pi_1(\mathcal{X}_{N,1,3},\hat{x}_{1 \cdots N})$ and $S_N$. Correspondingly, there is a short exact sequence
\begin{equation}
  1 \to PT_N \into T_N \twoheadrightarrow S_N \to 1.
\end{equation}

\end{document}